\documentclass[12pt]{article}

\usepackage{pic02}
\usepackage{hyperref}
\usepackage{url}
\usepackage{graphicx}

\def\Journal#1#2#3#4{{#1} {\bf #2}, #3 (#4)}

\def\EPC{{\em Eur. Phys. J.} C}
\def\CPC{Comp. Phys. Comm.}

\def\be{\begin{equation}}  
\def\ee{\end{equation}}
\def\bea{\begin{eqnarray}}
\def\eea{\end{eqnarray}}

\begin{document}
\title{\bf PRECISION TESTS OF QCD AT HERA}
\author{
David Milstead        \\
{\em Department of Physics, University of Liverpool, Liverpool, UK.}}
\maketitle

%
%
%
%
%
%
\vspace{4.5cm}
%

\baselineskip=14.5pt
\begin{abstract}

The electron-proton collider HERA has allowed the study of the
partonic content of the proton in regions of $Q^2$ up to 50,000 GeV$^2$
and values of Bjorken-$x$ down to $10^{-5}$.
This paper presents recent precision measurements of
structure functions and hadronic final state observables which test QCD 
over this wide region of phase space.

\end{abstract}
\newpage

\baselineskip=17pt

\section{Introduction}

Deep-inelastic lepton-nucleon scattering experiments have traditionally
shed light on the nature of the partons within the proton and the
strong QCD interactions between them\cite{qcdt}.

The $ep$ scattering experiments H1 and ZEUS at the HERA
facility at DESY have continued this tradition and have made precision
measurements of both the inclusive 
$ep$ scattering cross-section and features of the $ep$-induced hadronic
final state.
The theory of perturbative QCD, as implemented in the
 DGLAP\cite{DGLAP} parton evolution equations, has been successful in
describing these
measurement at higher values of $Q^2$  (above $\approx$ 100 GeV$^2$).
However, the DGLAP equations resum terms in $ln(Q^2)$ and ignore terms in
$ln(1/x)$ which are expected to become important in the lower $Q^2$, low
$x$ domain ($x\leq 10^{-3}$)\cite{BFKL}. Furthermore, gluon recombination
effects may also be visible in the low $x$ domain\cite{gr}.

In a complementary way, the hadronic final state from $ep$ collisions also offers
a rich testing ground of QCD. High transverse momentum jet
and particle production particles directly probe pQCD processes and
reveal the partonic structure of the proton and photon. Hadron
properties at low momentum are sensitive to the hadronisation process
and can thus be used to assess the environmental dependence
of fragmentation and test non-perturbative QCD models of hadron production.

This paper presents precision measurements of the proton
structure $F_2$ and of hadronic final state observables.

\section{The Proton Structure Function $F_2$ and the Gluon Momentum
Distribution $xg(x)$ of the 
Proton}

The reduced $ep$ scattering cross-section $\sigma_r$ can be written as a
function of $F_2$ and $F_L$.
$$\sigma_r={d\sigma \over dxdQ^2}\cdot{Q^4x \over 2\pi \alpha^2
Y_+}=F_2(x,Q^2)- {y^2F_L(x,Q^2) \over Y+}$$

This quantity is dominated by the contribution from $F_2$ and
$F_L$ only plays an important role at large values of the
inelasticity $y={Q^2\over sx}$.

\begin{center}
\begin{figure}[htb]
\begin{center}
\includegraphics[width=11cm]{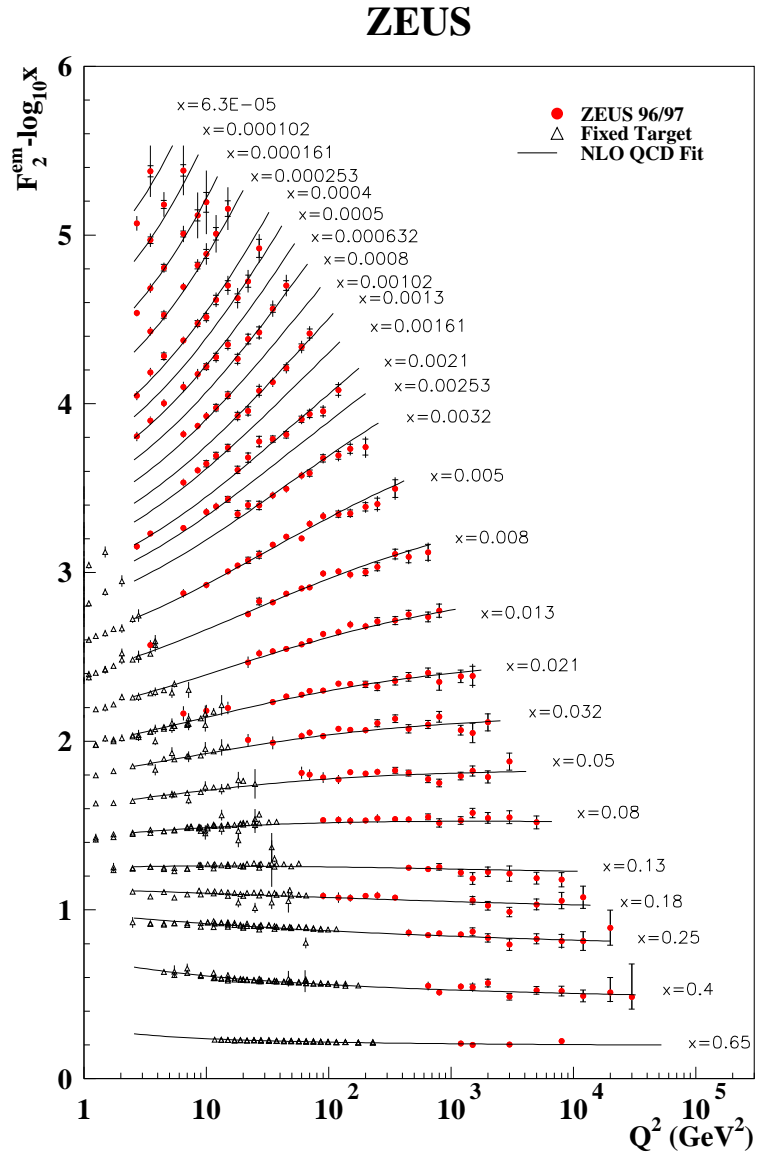}
 \caption{\it
The proton structure function $F_2(x,Q^2)$ shown as a function of
$Q^2$ for fixed values of $x$. QCD fits and results from fixed target
experiments are also shown.} \label{fig:fig1}
\end{center}
\end{figure}
\end{center}

Precision measurements of $F_2(x,Q^2)$ have been made by the
H1\cite{h1fullf2} and ZEUS\cite{zeusfullf2} collaborations. The ZEUS
results are  shown in figure~\ref{fig:fig1} and compared with earlier 
fixed target data.
The classic
scaling violations of $F_2$ are exhibited via the strong dependence on 
$Q^2$ for a range of fixed $x$ values.  The low $x$ behaviour of $F_2$ is
dominated by  quark anti-quark pair
production arising from the gluon content $g(x)$ of the proton
which, within the DGLAP formalism, is given by $F_2 \propto \alpha_s\cdot
g(x)$. QCD fits based on DGLAP evolution  using the HERA data 
and the fixed target data are also shown. The 
fit describes  the data well over several orders of magnitude in $x$ and
$Q^2$.  In the region of kinematic overlap the independent 
measurements from fixed target experiments and the ZEUS data agree well.

\begin{figure}[!htb]
\begin{center}
\includegraphics[width=11cm]{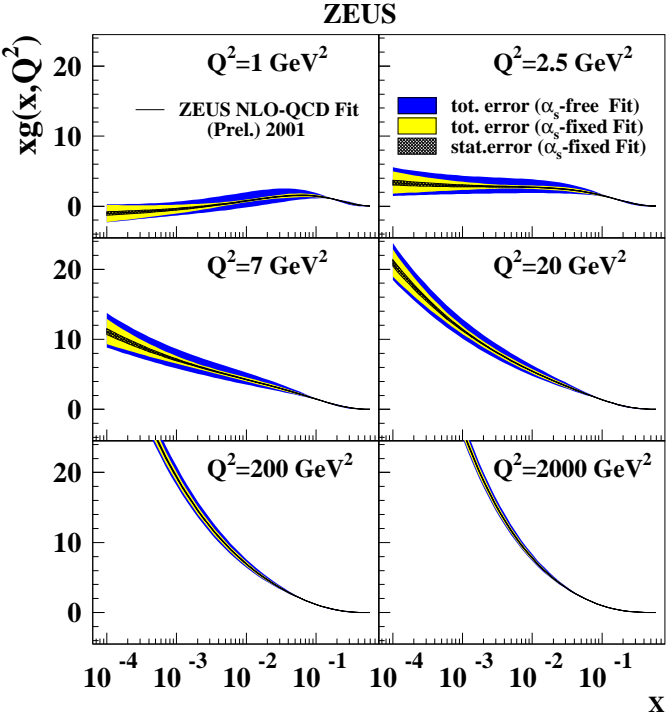}
\caption{The gluon momentum distribution $xg(x)$ shown as a function of
$x$ 
at different values of $Q^2$. }
\label{fig:fig2}
\end{center}
\end{figure}

Using these precision measurements of the $ep$ scattering cross-section
and within the DGLAP formalism the gluon momentum distribution $xg(x)$ has
been derived\cite{zeusprel}. 
The dependence of $xg(x)$ on $x$ for different values of $Q^2$ is shown in
figure~\ref{fig:fig2}. There is a strong rise towards low $x$ for values
of $Q^2$ down to about 5 GeV$^2$. Below this, the gluon distribution tends
to flatten off, approaching valence quark-like behaviour.   

\section{The Longitudinal Structure Function $F_L$} 
\begin{figure}[!htb]
\begin{center}
\includegraphics[width=11cm]{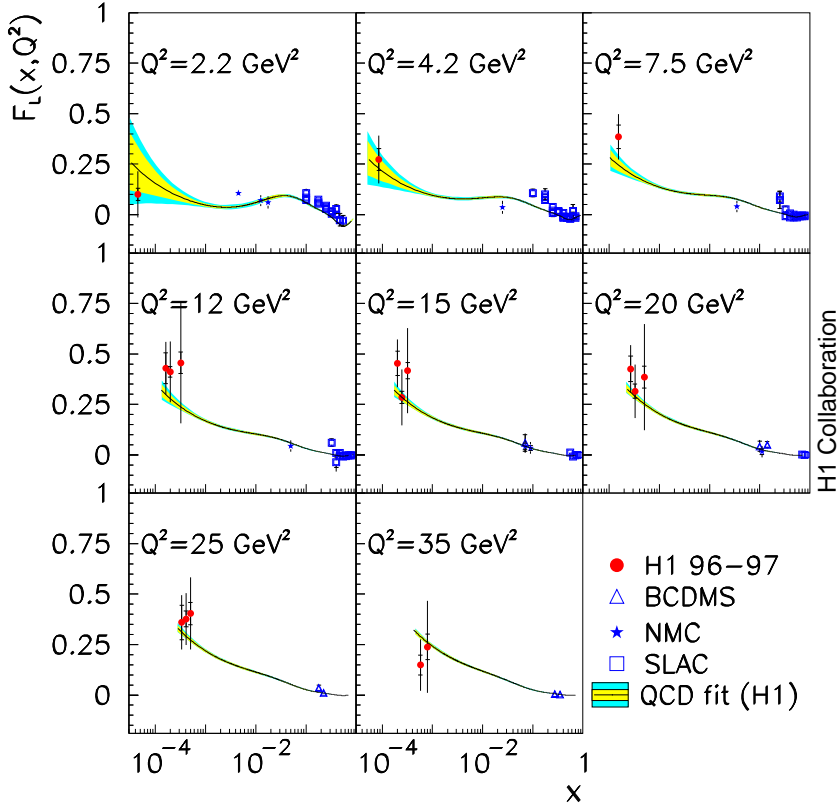}
\caption{The longitudinal structure function $F_L$ shown as a function
of $x$ at different values of $Q^2$. }
\label{fig:fig3}
\end{center}
\end{figure}

At high values of $y$ the contribution from the
longitudinal structure function $F_L$ becomes significant. 
In figure~\ref{fig:fig3}, measurements of $F_L$ from the H1 collaboration 
are presented as a
function
of $x$ at different values of $Q^2$.

The values of $F_L(x,Q^2)$ increase towards low $x$, consistent with the
QCD calculation which reflects the rise of the gluon momentum
distribution.

The values of $F_L$ were experimentally determined for $Q^2<10$
GeV$^2$ following
a procedure involving fitting straight lines in $ln y$ to the derivative
${\partial \sigma_r \over \partial ln y}$ in order to estimate the
contribution from the behaviour of $F_2$ to the reduced cross-section. 
For $Q^2>10$ GeV$^2$, the NLO QCD fit in the low $y$ domain ($y<0.35$) is
used to estimate $F_2$ in the high $y$ region and thus extract $F_L$. 

\section{The Rise of $F_2$ Towards Low $x$}

As mentioned in the introduction, the low $x$ dependence of $F_2$ may be
directly sensitive to any new
physics which may be evident in this region of high partonic
density\cite{gr}. 
  
Fits of the form $F_2(x,Q^2)=c(Q^2)x^{-\lambda(Q^2)}$ were made to the 
H1 structure function data\cite{h1rise}.

\begin{figure}[!htb]
\begin{center}  
\hspace{-1.cm}
\includegraphics[width=11cm]{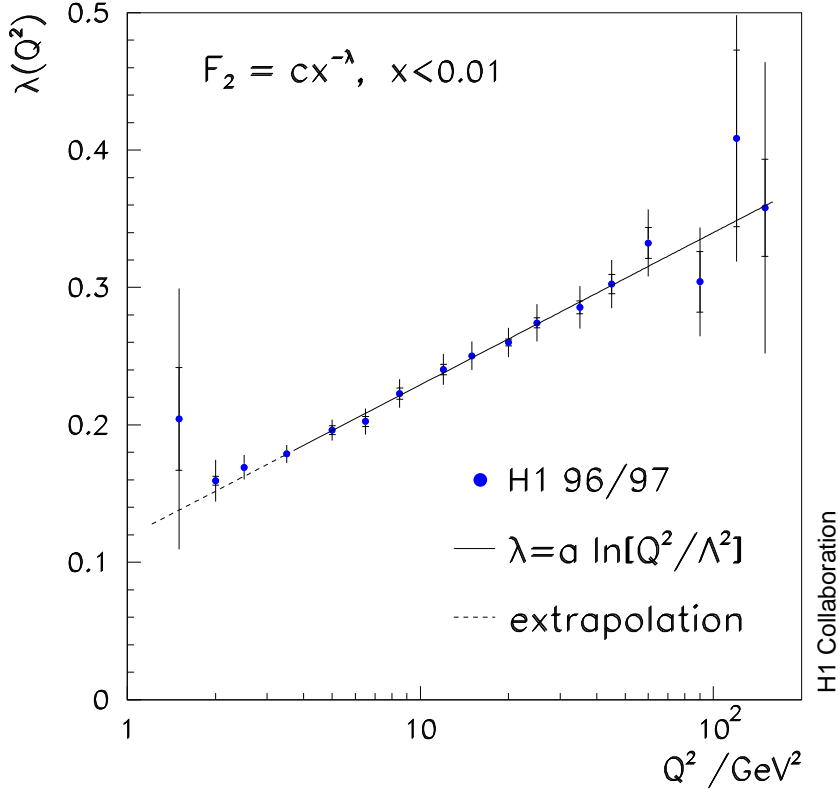}
%
\caption{The dependence of $\lambda$ on  $Q^2$. }
\label{fig:fig4}
\end{center}
\end{figure}

The co-efficients $c(Q^2)$ were found to be approximately independent of
$Q^2$ and, as shown in figure~\ref{fig:fig4}, $\lambda(Q^2)$  rises
approximately linearly with $ln Q^2$ and can be represented as
$\lambda(Q^2)=a\cdot ln[{Q^2 \over \Lambda^2}]$ with co-efficients
$a=0.0481\pm 0.0013(stat) \pm 0.0037(syst)$ and $\Lambda=292\pm
20(stat)\pm 51(syst)$ MeV.

Below the deep-inelastic scattering region, for fixed $Q^2<1$ GeV$^2$,
the simplest Regge phenomenology predicts that $F_2(x,Q^2)\propto
x^{-\lambda}$ where $\lambda=\alpha_{pom(0)}-1 \approx 0.08$\cite{reg}.
An extrapolation of the function $\lambda(Q^2)$ from the QCD fit gives a
value of 0.08 at $Q^2=0.45$ GeV$^2$.

\begin{center}
   \begin{figure}[ht!] \unitlength 1cm
      \begin{center}  
\vspace{0.cm}
\includegraphics[width=11cm]{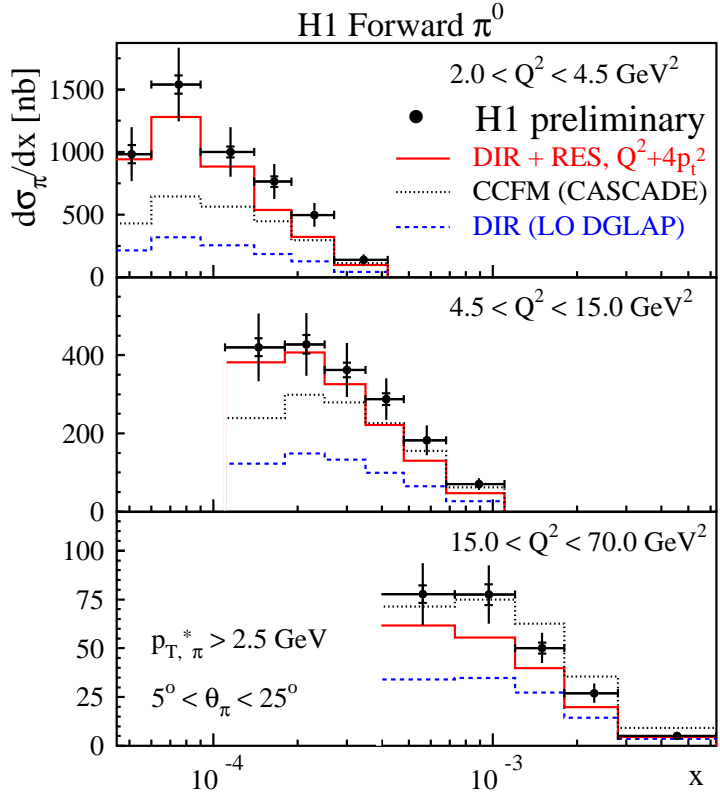}
      \end{center}  
\vspace{0.3cm}
\caption{The forward $\pi^0$ cross-section
as  a function of Bjorken-$x$ for different intervals of $Q^2$.}
\end{figure}
\label{fig:fpi}
\end{center}

\section{Forward $\pi^0$ Production}

The presence of particles with large values of transverse momentum
($p_T$) is indicative of a hard partonic sub-process
and thus can be used to
directly test pQCD calculations. It was therefore
suggested that
measurements of the forward $\pi^0$ cross-section could help to resolve
one of the open
questions at HERA: how do partons evolve from the proton ? Parton
evolution based on the
DGLAP~\cite{DGLAP}
equations successfully describes all inclusive measurements of the proton
structure function $F_2$. However,
this scheme sums terms in $ln(Q^2)$ and ignores terms in $ln(1/x)$ which
should play a significant
role in the low Bjorken-$x$ domain opened by HERA and therefore
BFKL~\cite{BFKL} evolution which,
conversely, sums terms in $ln(1/x)$
is expected to become applicable. Although measurements of $F_2$ offer no
discrimination between the two schemes,
the different dynamical features of the two approaches suggests that the
hadronic final state may offer a means
of observing BFKL effects. DGLAP evolution imposes a strong increasing
ordering on $k_T$, the transverse momentum of 
the emitted parton and a weak decreasing ordering in the fractional
momentum $x$ of the propagating parton as the
``ladder" develops away from the proton.  However,
BFKL evolution imposes a strong ordering in $x$ and no ordering in $k_T$,
allowing
partons to take a "random walk" in $k_T$ and thus
would give rise to more high momentum parton emissions than from DGLAP
evolution.

It would then be expected that
measurements of particle production in the region associated with this
parton ladder would be sensitive to evolution
effects.
In fact,  a series of hadronic final state
measurements on transverse energy flow and jet and particle production in
the forward (ie close to the proton)
direction have already been made. A recent
highlight of these works is the
measurement of high $p_T$ $\pi^0$ mesons~\cite{tw}.
These are made for $\pi^0$ mesons with values of transverse momentum in
the photon-proton hadronic centre of
mass system of $p_T>2.5$ GeV. Values of the polar angle, $\theta$,
measured with respect to the incoming proton
 direction in the 
 laboratory system were restricted to $5^0<\theta<25^0$ and the laboratory
system $\pi^0$ energy scaled by the
incoming proton energy
$(x_\pi =
E_\pi/E_{proton})$ was required to be greater than 0.01. The deep
inelastic scattering
(DIS) kinematic range in $Q^2$ and the inelasticity variable, $y$, was
restricted to
 $2 <Q^2 < 70$ GeV$^2$ and $0.1<y<0.6$.

Figure~\ref{fig:fpi} shows the forward $\pi^0$ cross-section  and
 as a function of Bjorken-$x$ for three
different intervals of $Q^2$. The cross-section rises towards low
Bjorken-$x$ for each of the $Q^2$ ranges. 
 Calculations based on DGLAP evolution from the
proton alone fail to describe the data. Partonic
structure can also be
 assigned to the resolved photon thus allowing DGLAP evolution from the
photon
and the proton. Calculations based on this picture are also shown~\cite{rapgap}
 and
they come close to the data although these fail at the highest measured
of $Q^2$. Calculations based on the CCFM equation which should interpolate
between the DGLAP and BFKL regimes describes the data only in the highest
 $Q^2$ interval.

It is the author's opinion that the next step for these analyses must be
the exploitation of the full
$ep$ event information. One such proposed analysis is using multi-particle
correlations which are directly sensitive to
 the  presence or absence of strong $k_T$ ordering in parton
emissions~\cite{eddi}. The H1 and ZEUS experiments are strongly
encouraged to pursue this.

\section{Summary}

Results on deep-inelastic $ep$ scattering cross-sections in the low $x$
regime at HERA have been presented. Perturbative QCD calculations
based on DGLAP evolution have been tested against precision results on
proton structure functions and found to describe the data over many orders
of magnitude in the kinematic variables $x$ and $Q^2$.



\begin{thebibliography}{99}
\bibitem{qcdt} See, for example, M.~Klein, {\em Structure Functions in
Deep-Inelastic Lepton-Nucleon Scattering}, Proc. Lepton-Photon Symposium,
Stanford, August 1999, World Scientific, ed. by J. Jaros and M. Peskin, p.
467, hep-ex/0001059 (2000).

\bibitem{DGLAP} Yu.~L.~Dokshitzer, Sov. Phys. JETP {\bf 46} (1977) 641;\\
V.~N.~Gribov and L.~N.~Lipatov, Sov. J. Nucl. Phys. {\bf 15} (1972) 438
and 675; \\
G.~Altarelli and G.~Parisi, Nucl. Phys. {\bf B126} (1977) 298. 

\bibitem{BFKL} E.~A.~Kuraev,L.~N.~Lipatov and V.~S.~Fadin, Sov. Phys.
JETP{\bf 44} (1976) 443; \\
 E.~A.~Kuraev,L.~N.~Lipatov and V.~S.~Fadin, Sov. Phys. 
JETP{\bf 45} (1977) 199; \\
Y.~Y.~Balitsky and L.~N.~Lipatov, Sov. J. Nucl. Phys. {\bf 28} (1978)
822.

\bibitem{gr} L.~V.~Gribov, E.~M.~Levin and M.~G.~Ryskin, Phys. Rep. {\bf
100} (1983) 1.
\bibitem{h1fullf2} H1 Collaboration, C. Adloff {\it et al.}, Eur. Phys. J.
{\bf C21} (2001)
33.
\bibitem{zeusfullf2} ZEUS Collaboration, S.~Chekanov {\it et al.}, Eur. Phys. J. {\bf C21}
(2001) 443.
\bibitem{zeusprel} ZEUS Collaboration, submitted paper 628, International
Europhysics Conference on High Energy Physics, Budapest, Hungary,
July12-18,2001.

\bibitem{h1rise}  H1 Collaboration, C.~Adloff {\it et al.}, Phys. Lett. {\bf B520} (2001)
183.

\bibitem{reg}     A.~Donnachie and P.~V.~Landshoff, Z. Phys. {\bf C61} (1994) 139.
\bibitem{tw} C.~Adloff et al., Phys. Lett. {\bf B462}{440}{1999}.
\bibitem{bfklmart}  J.~Kwiecinski, A.~Martin and 
J.J.~Outhwaite \Journal{\EPC}{9}{1999}{611}.
\bibitem{lepto} G. Ingelman,
                Proc. of the HERA workshop, eds W.~Buchm\"uller and
                G.~Ingelman, Hamburg (1992) Vol. 3, 1366.
\bibitem{rapgap} {H. Jung}, \Journal{\CPC}{86}{147}{1995}
(for update see http://www-h1.desy.de/\string ~jung/rapgap/rapgap.html).
\bibitem{eddi} E.~de~Wolf and P. Van~Mechelen, Proc. of the Workshop on
Monte Carlo Generators for HERA Physics, 1999, eds A.T.~Doyle,
G.~Grindhmmer,
G.~Ingelman and H.~Jung.

\end{thebibliography}
\end{document}